\documentclass[11pt]{article} 

\usepackage[utf8]{inputenc} 
\usepackage{amsmath,amssymb,multirow,multicol}

\usepackage{geometry} 
\usepackage{graphicx} 

\usepackage{booktabs} 
\usepackage{array} 
\usepackage{paralist} 
\usepackage{verbatim} 
\usepackage{subfig} 

\usepackage{fancyhdr} 
\usepackage{sectsty}
\allsectionsfont{\sffamily\mdseries\upshape} 

\usepackage[nottoc,notlof,notlot]{tocbibind} 
\usepackage[titles,subfigure]{tocloft} 
\usepackage{makecell} 

\allowdisplaybreaks 

\newcommand{\etal}{\emph{et al.}}
\newcommand{\tZ}{\tilde{Z}}
\newcommand{\cor}{\mathrm{Cor}}
\newcommand{\cov}{\mathrm{Cov}}

\newcommand{\var}{\mathrm{var}}
\newcommand{\pixi}{\pi_{X,i}}
\newcommand{\pix}{\pi_{X}}
\newcommand{\pixx}{\pi_{XX}}
\newcommand{\piz}{\pi_{Z}}
\newcommand{\VL}{V_{\mathrm{LCRT}}}
\newcommand{\sigzz}{\sigma^2_Z}
\newcommand{\sigxx}{\sigma^2_X}
\newcommand{\bb}{\sigma_{CT}^2}

\newcommand{\xx}{\mathbf{q}}
\newcommand{\dd}{\mathbf{d}}
\newcommand{\sve}{\sigma_\varepsilon^2}

\newcommand{\rhoC}{\rho_{\mathrm{C}}}
\newcommand{\rhoCT}{\rho_{\mathrm{CT}}}
\newcommand{\rct}{r_{\mathrm{CT}}}
\newcommand{\betaI}{\beta_\mathrm{I}}
\newcommand{\betaC}{\beta_\mathrm{C}}
\newcommand{\betaCI}{\beta_\mathrm{IC}}
\newcommand{\tbetaC}{\tilde{\beta}_{\mathrm{C}}}
\newcommand{\betaj}[1]{\beta_{#1}}
\newcommand{\tbeta}{\tilde{\beta}}
\newcommand{\tbetaj}[1]{\tilde{\beta}_{#1}}
\newcommand{\hbetaI}{\hat\beta_\mathrm{I}}
\newcommand{\hbetaC}{\hat\beta_\mathrm{C}}
\newcommand{\hbetaCI}{\hat\beta_\mathrm{IC}}
\newcommand{\htbetaC}{\hat{\tilde{\beta}}_{\mathrm{C}}}

\newcommand{\hBetaT}{\hat{\tilde{\boldsymbol{\beta}}}}

\title{Sample size calculations for multilevel factorial longitudinal cluster randomised trials}
\author{Rhys Bowden, Rebecca Walwyn, Jessica Kasza, Andrew Copas, Fan Li, \\James Wason, Andrew Forbes}
\date{} 

\begin{document}
\maketitle

\begin{abstract}{Typically, trials investigate the impact of either an individual-level intervention on participant outcomes, or the impact of a cluster-level intervention on participant outcomes. Factorial designs consider two (or more) treatments for each of two (or more) different factors. In factorial trial designs, trial units (individuals or clusters) are each randomised to a level of each of the treatments; these designs allow assessment of the interactions between different interventions. Recently, there has been growing interest in the design of trials that jointly assess the impact of individual- and cluster-level interventions (i.e. multi-level interventions); requiring the development of methodology that accommodates randomisation at multiple levels. While recent work has developed sample size methodology for variants combining standard cluster randomisation and individual randomisation, that work does not apply to longitudinal cluster randomised trial designs such as the stepped wedge design or cluster randomised crossover design. Here we present dedicated sample size methodology for “split-plot factorial longitudinal cluster randomised trials” with continuous outcomes: allowing for joint assessment of individual-level and cluster-level interventions that allows for the impact of the cluster-level intervention to be assessed using any longitudinal cluster randomised trial design. We show how the power to detect given effects of the individual-level intervention, the cluster-level intervention, and the interaction between the two depends on standard results for individually-randomised trials and longitudinal cluster randomised trials. We apply these results to the SharES trial, which considered the effects of a patient- and clinician-level interventions for patients with breast cancer on patient knowledge about the risks and benefits of treatment. 
}
\end{abstract}

\section{Introduction}
Trial time and data are valuable: testing the effect of an intervention often requires the enrolment of a large number of participants, or a substantial number of institutions or treatment centres. 
This is particularly true in the case of cluster-level interventions that may need to enrol a high proportion of the available clusters in a country or jurisdiction in order to effectively measure the impact of an intervention. However, it is often the case that there are a large number of interventions to differing aspects of treatment that can be randomised independently but need to be evaluated for their effect on the same outcome (like mortality rate or time spent in hospital); so it is of great benefit to be able to evaluate multiple interventions simultaneously on a single population of participants. This is especially the case in areas like intensive care: where there are a wide variety of possible competing treatments for different aspects of patient care, some of which are impractical or impossible to implement on an individual level. In many cases, there is also interest in the interaction between these different interventions, or in finding the best combination of treatments.

This gives rise to factorial trials: trials that allow for separate and independent randomisation of treatment factors. In the case of cluster randomised factorial trials, all the factors may be randomised individually, or some factors may be randomised individually and others randomised at a cluster-level.  There is increasing interest in trials with multi-level interventions~\cite{mcbride2021walking}, including those that combine individual and cluster-level interventions~\cite{mdege20142,nahum2018multilevel}, particularly in application areas that wish to test behavioural interventions; in many cases smartphone applications or supplied education material allow delivery of an individually randomised treatment that may supplement or reinforce the effect of an intervention that is delivered at the cluster-level~\cite{kerr2018cluster}. Most commonly, these sorts of trial use the so-called \emph{split-plot} design, where clusters are randomised to a treatment condition, and separately, the individuals within each cluster are randomised separately, in such a way that each cluster has a similar proportion of participants randomised to each individual-level treatment condition.

The theory for individually-randomised factorial trials is very well-established, but some of the theory for trials that mix individual and cluster-randomisation has only recently been solidified~\cite{tian2022sample}; and even more so for \emph{split-plot longitudinal cluster randomised trials}: trials that mix individual-level interventions with longitudinal cluster randomised trials (LCRTs), where each cluster is randomised to a \emph{sequence} of treatments. A well-known and increasingly popular example of a longitudinal cluster randomised trial is the stepped-wedge design: a design where all clusters start in the control condition and the intervention is gradually rolled out over the clusters so that each cluster finishes the trial in the intervention condition. Stepped wedge trials are particularly useful for interventions that cannot be undone (like educational or training interventions), or interventions that will be rolled out anyway but there is interest in measuring their effectiveness.

This paper considers not just stepped wedge trials, but LCRTs in general, and is the first to derive closed-form expressions for the required sample size for factorial trials that have an individually randomised treatment intervention, and an intervention randomised in an LCRT design (such as stepped wedge). Further, these expressions are derived for any given split-plot LCRT with continuous outcomes in terms of the variance expressions for the simple single-level LCRT (the trial with no individually randomised component). They allow for both the situation where an interaction between the different interventions is included in the model, and the situation where there is assumed to be no interaction. Calculating sample size is a crucial task for any researchers who wish to conduct a factorial LCRT, and expressions for sample size enable understanding of the key components contributing to sample size; while also facillitating purposeful design and quick evaluation of competing potential trial designs.

The remainder of the paper is divided into a background section (Section 2), where information about the existing literature on factorial and factorial cluster trials is given, and the statistical model for the outcome is provided; Section 3 that includes the variance expressions for the treatment effect estimators in the scenarios where the statistical model includes an interaction term or not; Section 4 for how to use these expressions to calculate the sample size in practice; Section 5 considers examples of these sample size calculations applied to a real trial; and Section 6, a discussion and conclusions.

\section{Background}
\subsection{Existing literature}
Factorial designs have a long history in both agricultural research and engineering~\cite{mukerjee2007modern}. Historically, factorial multi-level trials or experiments whereby one unit of randomisation is nested within another unit of randomisation are often called split-plot designs, due to their origins within the agricultural testing literature. In this literature, there would be large plots split into smaller subplots, and the large plots would be randomised to one particular large-plot-level treatment, while the smaller plots would each be randomised to a second small-plot-level treatment; outcome measurements would be taken on small plots or individual plants. In the clinical trials world, an obvious analogy are trials where clusters of participants are randomised to a cluster-level treatment, while individuals within those clusters are randomised separately and independently to an individual-level treatment. 

Over the past two decades or so there have been many factorial cluster randomised trials conducted in healthcare~\cite{mdege20142}; some of these have all factors randomised at the cluster level, but many of them are split-plot designs with at least one factor randomised at the individual-level. If it is possible to individually-randomise a factor then this can be strongly advantageous in terms of power and precision: a well-known result states that typically both the effect of the individual level intervention \emph{and} its interactions with other intervention effects can be estimated with a variance that scales with the number of individuals in the study, rather than with the number of clusters. That is, intervention effects are then only subject to individual-level variability, not cluster-level variability~\cite[Section 19.1]{christensen2018analysis}. More recent methodological work on cluster-randomised split-plot designs has focused on advanced methods such as sample size calculations when cluster sizes are variable~\cite{tian2022sample}.

Despite the variety of work for standard (parallel) cluster randomised factorial designs~\cite{dziak2012multilevel}, there is little information and few resources on split-plot designs for longitudinal cluster randomised trials like stepped-wedge designs.  A notable exception to this is Sperger \etal~\cite{sperger2024multilevel} which provides an introduction and definition of multi-level stepped wedge designs, and proposes using simulation to calculate sample sizes for these designs. Most notably, Sperger \etal{} doesn't provide analytical formulas for sample size that are provided here. Having analytical expressions for sample size provides at least two main benefits: first, it may allow for faster consideration of a range of scenarios when assessing differing design and parameter choices; and second, it provides a better understanding of the sample size, and how each component of the sample size contributes to the overall result.

Recent interest in these sorts of designs is increasing, driven by the increased ability and desire to provide interventions on multiple levels, and also the increased popularity of hybrid implementation/effectiveness designs. Sometimes factorial stepped wedge trials with some factors that might be possible to randomise at the individual level have all their factors randomised at the cluster level instead~\cite{van2020learning,rutten2024multilevel}, foregoing the benefits of individual level randomisation, perhaps due to paucity of resources for stepped wedge split-plot designs. Nevertheless, there are already split-plot stepped wedge designs being used in practice, we apply our methods to one in particular~\cite{hawley2023improving} as a worked example in Section \ref{sec:example}.

\subsection{Model}

Let the observation $Y_{i,j,k}$ for cluster $i=1,\ldots,n$, period $j=1,\ldots,T$, individual $k=1,\ldots,m_{i,j}$ be
\begin{align}
Y_{i,j,k} &= \betaC X_{i,j} + \betaI Z_{i,j,k} + \betaCI X_{i,j}Z_{i,j,k} + \betaj{j} + a_i + b_{i,j} + \varepsilon_{i,j,k}
\end{align}
where $a_i$ are independent and identically distributed with $a_i \sim N(0,\sigma_a^2)$, $b_{i,j}$ are independent and identically distributed with $b_{i,j}\sim N(0,\sigma_b^2)$, and $\varepsilon_{i,j,k}$ are independent and identically distributed with $\varepsilon_{i,j,k} \sim N(0,\sve)$. $X_{i,j}$ is the cluster-period-level treatment indicator and $Z_{i,j,k}$ is the individual-level treatment indicator (both 0-1 variables that are 1 to signify allocation to the intervention, and 0 to signify allocation to the control for their respective treatments). $\betaj{j}$ is the fixed effect for time in period $j$, and $\betaC$, $\betaI$ and $\betaCI$ are the targets of inference. $\betaC$ is the effect of the cluster-level treatment when the individual-level treatment is in the control condition, $\betaI$ is the effect of the individual-level treatment when the cluster-level treatment is in the control condition, and $\betaCI$ is the interaction between the effects of the cluster- and individual-level treatments. 

The individual-level intervention is assumed to be assigned randomly within each cluster-period, so that $Z_{ijk}$ is independent of $X_{ij}$ and $i$ and $j$, and such that of the $m$ observations in each cluster period, $\piz m$ have $Z_{ijk}=1$ and the remainder have $Z_{ijk}=0$. Further, let $N=\sum_{i=1}^n\sum_{j=1}^T m_{i,j}$, the total number of observations, and let
\begin{align}
\pi_{X,i} &=\frac{1}{\sum_{j=1}^Tm_{i,j}}\sum_{j=1}^T m_{i,j} X_{ij}\mbox{, and }\\
\pi_X &= \frac{1}{N}\sum_{i=1}^n\sum_{j=1}^T m_{i,j} X_{ij},
\end{align}
that is, $\pixi$ is the proportion of observations assigned to the cluster-level intervention of those in the $i$th cluster, and $\pix$ is the proportion of observations assigned to the cluster-level intervention overall. Also, $\piz$ is the proportion of individuals assigned to the individual-level intervention, both overall and in each period.

The cluster and cluster-period random effects are $a_i,b_{ij}$ and the residual is $\varepsilon_{ijk}$, which together encode a block exchangeable correlation structure. This correlation structure has a separate between-period and within-period correlation, denoted respectively by
\begin{align}
 \rhoC = \cor(Y_{ijk_1},Y_{ijk_2}) = \frac{\sigma_a^2}{\sigma_a^2+\sigma_b^2+\sve} \\
\rhoCT = \cor(Y_{ij_1k_1},Y_{ij_2k_2}) = \frac{\sigma_a^2}{\sigma_a^2+\sigma_b^2+\sigma_b^2+\sve} .
\end{align}
for $k_1,k_2 = 1,\ldots,m$; $j_1,j_2 = 1,\ldots,T$ with $j_1 \neq j_2$ and $k_1 \neq k_2$ and $\rhoC\leq \rhoCT$. Individuals are assumed to contribute measurements to one period and cluster only (cross-sectional recruitment).

For ease of the following derivations we also reparametrise the model with $\tZ_{ijk} = Z_{ijk}-\piz$ as follows:
\begin{align}
Y_{ijk} &= \betaC X_{ij} + \betaI Z_{ijk} + \betaCI X_{ij}Z_{ijk} + \betaj{j} + a_i + b_{i,j} + \varepsilon_{ijk}\\
&= (\betaC+\pi_Z\betaCI) X_{ij} + \betaI(Z_{ijk}-\pi_Z) + \betaCI X_{ij}(Z_{ijk}-\pi_Z) + (\betaj{j}+\pi_Z\betaI) + a_i + b_{ij} + \varepsilon_{ijk}\\
&= \tbetaC X_{ij} + \betaI\tZ_{ijk}+ \betaCI X_{ij}\tZ_{ijk} + \tbeta_{j} + a_i + b_{ij} + \varepsilon_{ijk}.
\end{align}
Here $\tbetaC$ is the average effect of the cluster-level intervention when the probability of being given the individual-level intervention is fixed at $\pi_Z$; this is often referred to as the ``at the margins'' estimate as it relates to the margins of the 2 by 2 table of possible treatment groups, or the \emph{marginal effect} of the cluster-level treatment when the population is exposed to the individual-level intervention in the same proportions as they are in the trial. The original interaction effect and time effects are given in terms of the reparameterised interaction effect and time effects by
\begin{align}
\betaC &= \tbetaC-\betaCI\pi_Z\\
\betaj{j} &=\tbetaj{j}-\betaI \pi_Z.
\end{align}

\section{Variance of treatment effect estimators}
\subsection{Including an interaction term in the model}
We start by considering the case where an interaction term $\betaCI$ is included in the model. For the estimation of sample size, a vital component is the variance of the treatment effect estimator. In the case of multiple treatment efffects as we have here, the key matrix is the asymptotic (equivalent to feasible generalized least squares (FGLS) or generalized least squares (GLS) estimators conditional on the assumed variance/correlation parameters) covariance matrix of the effect estimators $\hbetaC,\hbetaI,\hbetaCI$. In order to find this matrix it is simpler to start by finding the covariance matrix of $\htbetaC,\hbetaI,\hbetaCI$ then derive the desired covariances with $\hbetaC$. 

In Appendices \ref{sec:2} and \ref{sec:var_size} we derive the covariance matrix of the treatment effect estimators $\hBetaT=(\hbetaI,\hbetaCI,\htbetaC)$, it is given by:
\begin{align}
\var(\hBetaT) = \begin{bmatrix}
K\pix & -K\pix & 0\\
-K\pix & K & 0\\
0 & 0 & \VL
\end{bmatrix}
\end{align}
where $K=\frac{(1-\rhoCT)\sigma^2}{\sigzz \sigxx N}$ and $\sigzz = \piz (1-\piz)$, $\sigxx=\pix(1-\pix)$ and $\VL$ is the variance of the standard treatment effect estimator for the longitudinal cluster randomised trial of the same design but with only a cluster-period-level intervention (\emph{i.e.} if we ignored the individual-level treatment entirely). Formulas for $\VL$ are available in LCRT literature depending on the design of the trial and other features of the mathematical model, we give an example one in Section \ref{sec:sample_size}. We can now pick out the variance of the individual $\beta$ parameter estimators from the diagonal elements of $\var(\hBetaT)$, to give:
\begin{align}
\var(\hbetaI) &= \frac{(1-\rhoCT)\sigma^2}{\sigzz \pix N} = \frac{(1-\rhoCT)\sigma^2}{\sigzz N_{X=0}},\\
\var(\hbetaCI) &= \frac{(1-\rhoCT)\sigma^2}{\sigzz \sigxx N} = \frac{(1-\rhoCT)\sigma^2}{\sigzz}\left(\frac{1}{N_{X=1}} + \frac{1}{N_{X=0}}\right)\mbox{, and}\\
\var(\htbetaC) &= \VL \label{eq:beta2_var_intM}
\end{align}
To get $\var(\hat\betaC)$ recall that $\betaC = \tbetaC-\pi_Z\betaCI$ so
\begin{align}
\var(\hbetaC) &= \var(\htbetaC-\hbetaCI\piz)\notag\\
&=\var(\htbetaC)+\piz^2\var(\hbetaCI)-2\cov(\htbetaC,\hbetaCI\piz)\notag\\
&=\VL + \frac{\piz^2(1-\rhoCT)\sigma^2}{\sigzz\sigxx N}.\label{eq:beta2_intM}
\end{align}

\subsection{Without an interaction term in the model}
It is also possible to derive variance expressions when there is assumed to be no interaction between the different factors. In this case there is no interaction term ($\betaCI$ or $\tbeta_4$) in the model, but otherwise the situation is identical. Appedices \ref{sec:no_int} and \ref{sec:var_size_no_int} give the expressions for the variances of the two treatment effect estimators:
\begin{align}
\var(\hbetaC) =  \VL\\
\var(\hbetaI) =\frac{(1-\rhoCT)\sigma^2}{\sigma_Z^2 N}
\end{align}

\begin{table}[ht]
\begin{center}
\caption{Expressions for the variance of the treatment effect estimator $\var(\hat\theta)$ under differing models and choices of $\theta$.\label{tab:summary}}
\begin{tabular}{p{2.5cm} p{1.5cm} p{6cm} p{5cm} } 
\hline
\textbf{Interaction included} & \textbf{$m$ varies} &\textbf{$\theta$} & \textbf{$\var(\hat\theta)$} \\
  \hline
Yes & No &Cluster-level intervention,      $\betaC$ & $\VL + \frac{\piz^2(1-\rhoCT)}{mT\sigzz\pix(1-\pix)}\cdot\frac{\sigma^2}{n}$\\ 
  \hline
Yes & No &Individual-level intervention,  $\betaI$ & $\frac{(1-\rhoCT)}{mT\sigzz(1-\pix)}\cdot\frac{\sigma^2}{n}$ \\ 
  \hline
Yes & No &Interaction,                         $\betaCI$ &$\frac{(1-\rhoCT)}{mT\sigzz\pix(1-\pix)}\cdot\frac{\sigma^2}{n}$\\ 
  \hline
No & No &Cluster-level intervention,   $\betaC$ & $\VL$ \\ 
      \hline
No & No &Individual-level intervention, $\betaI$ & $ \frac{1-\rhoCT}{mT \sigma_Z^2}\cdot\frac{\sigma^2}{n}$\\ 
  \hline
  Yes & Yes &Cluster-level intervention,      $\betaC$ & $ \VL+\frac{(1-\rhoCT)\sigma^2 \pi_Z N}{(1-\pi_Z)N_{X=1}N_{X=0}}$\\ 
  \hline
Yes & Yes &Individual-level intervention,  $\betaI$ & $\frac{(1-\rhoCT)\sigma^2}{\pi_Z(1-\pi_Z)N_{X=0}}$ \\ 
  \hline
Yes & Yes &Interaction,                         $\betaCI$ &$\frac{(1-\rhoCT)\sigma^2 N}{\pi_Z(1-\pi_Z)N_{X=1}N_{X=0}}$\\ 
  \hline
No & Yes &Cluster-level intervention,   $\betaC$ & $\VL$ \\ 
      \hline
No & Yes &Individual-level intervention, $\betaI$ & $ \frac{(1-\rhoCT)\sigma^2}{\sigma_Z^2 N}$\\ 
  \hline
\end{tabular}
\end{center}
\end{table}

\section{Sample size calculations}\label{sec:sample_size}
\subsection{Fixed cluster period size}
The standard sample size formula for a two-sided test for detecting a difference of size $\delta$ with power $1-\beta$ and significance level $\alpha$ is given by 
\begin{align}\label{eq:ss}
1 = \var(\hat\theta)\left(\frac{z_{1-\alpha/2}+z_{1-\beta}}{\delta}\right)^2,
\end{align} 
where $z_{1-\beta}$ and $z_{1-\alpha/2}$ are quantiles of the normal distribution, and $\theta$ is the treatment effect to be estimated, \emph{i.e.} one of $\betaC,\tbetaC,\betaI,\betaCI$. Depending on the treatment effect of interest and the chosen model, the appropriate expression for $\var(\hat\theta)$ can be used from Appendices \ref{sec:2} to \ref{sec:var_size_no_int} or Table \ref{tab:summary}. This expression is typically rearranged to get the required value of one parameter (such as number of clusters, or number of observations per cluster-period, or power, or effect size) in terms of the other parameters. For instance, if power $1-\beta$ is required, (\ref{eq:ss}) can be rearranged to give:
\begin{align}
1-\beta = \Phi\left(\frac{|\delta|}{\sqrt{\var(\hat\theta)}}-z_{1-\alpha/2}\right),
\end{align}
where $\Phi$ is the distribution function of the normal distribution.

In order to calculate the required sample size when $\theta=\betaC$ we will usually need an expression for $\VL$ in terms of the design and correlation parameters from the relevant literature (see~\cite{Hooper2016}). However, if we simply wish to calculate the required number of clusters per treatment sequence, $n_S$, then we can use the relationship that $\VL$ is inversely proportional to $n_S$ and use any standard sample-size calculator for LCRTs to obtain $\VL(1)$, the variance of the treatment effect estimator when there is one cluster per sequence, and then the variance of the single-level LCRT with $n_S$ clusters per sequence $\VL(n_S)$ will be $\VL(1)/n_S$.

If this approach is not possible, then it is possible to use the general form of the variance of the treatment effect estimator for the block exchangeable correlation structure cross-sectional design is given by~\cite{Hooper2016}:
\begin{align}
\VL = \frac{\sigma^2}{nm}\cdot\frac{S^2(1-\rct)(1+(T-1)\rct)(1+(m-1)\rhoCT)}{(SB-E+(B^2+S(T+1)B-(T+1)E-SC)\rct)},
\end{align}
where $\rct=\frac{m\rhoC}{1+(m-1)\rhoCT}$, $S$ is the number of treatment sequences in the design, and $B,C,E$ are determined by the cluster-period design matrix $\mathbb{X}=[X_{i,j}]$ with
\begin{align}
B &= \frac{1}{n_S}\sum_{i=1}^n\sum_{j=1}^T X_{i,j}\\
C &= \frac{1}{n_S}\sum_{i=1}^n\left(\sum_{j=1}^T X_{i,j}\right)^2\\
E &= \sum_{j=1}^T \left(\frac{1}{n_S}\sum_{i=1}^n X_{i,j}\right)^2.
\end{align}
Since $B,C,E$ are in terms of the cluster-period level design matrix $\mathbb{X}$ which has a number of rows $n=n_SS$, the normalisation by $n_S$ in the formulas above actually makes $B,C,E$ independent of $n_S$; they only depend on the form of the design.

To give an example of how to do the sample size calculation, assume we wish to calculate the required number of clusters per sequence $n_S$ to estimate the effect of the cluster-period level intervention in the absence of the individual level intervention, \emph{i.e.} $\betaC$, using a model with interactions. We can use the expressions (\ref{eq:beta2_int}) and (\ref{eq:ss}), and the relationship $n=n_S S$ to get:
\begin{align}
1 &=\frac{\sigma^2}{n_SSm} \left(\frac{S^2(1-\rct)(1+(T-1)\rct)(1+(m-1)\rhoCT)}{SB-E+(B^2+S(T+1)B-(T+1)E-SC)\rct}+\frac{\piz^2(1-\rhoCT)}{T\sigzz\pix(1-\pix)}\right)\left(\frac{z_{1-\alpha/2}+z_{1-\beta}}{\delta}\right)^2
\end{align}
This expression can be rearranged to get the required value of $n_S$:
\begin{align*}
n_S = \lceil \frac{\sigma^2}{Sm} \left(\frac{S^2(1-\rct)(1+(T-1)\rct)(1+(m-1)\rhoCT)}{SB-E+(B^2+S(T+1)B-(T+1)E-SC)\rct}+\frac{\piz^2(1-\rhoCT)}{T\sigzz\pix(1-\pix)}\right)\left(\frac{z_{1-\alpha/2}+z_{1-\beta}}{\delta}\right)^2\rceil
\end{align*}
where $\lceil x\rceil$ is the ceiling function, which rounds $x$ up to the nearest integer. Substituting in the desired values of the other parameters will give a value for $n_S$.

\section{Example of calculations for an existing trial - the SharES trial}\label{sec:example}
The SharES trial (Hawley et al~\cite{hawley2023improving}) is an oncology trial with a multi-level intervention consisting of (1) a patient-facing breast cancer treatment decision tool, and (2) a clinician-facing dashboard that summarizes ongoing patient needs for review by clinicians and/or surgeons. The trial is a factorial design, where intervention (1) is individually randomised and intervention (2) follows a hybrid parallel/stepped wedge design. ``Hybrid'' in this case means a hybrid stepped-wedge/parallel design: some clusters are randomised to control or intervention for the duration of the study, whereas others cross from the control condition to the intervention condition.

The trial runs over six 12-week time periods. There are 25 clusters randomised in total, with 5 clusters allocated to each of the two constant (parallel) sequences and 3 clusters allocated to each of the 5 standard standard stepped wedge sequences. 

For calculating a required sample size, we start by assuming an exchangeable correlation structure to match the assumption in the protocol, and assume an ICC of 0.2 ($\rhoC=\rhoCT=0.2$), the upper end of the range of ICCs considered in the protocol. In the absence of an interaction between the two different interventions, the required sample size is given as follows: To detect an effect size of $0.35$ standard deviations for the cluster-level effect with a power of $80\%$ and a significance level of $0.05$, $4$ obervations in each cluster period are required (matching the required sample size given in the protocol paper). To detect the same effect size in the individual-level effect would require 2 observations in each cluster period. However, if an interaction between the two interventions was included in the model, then 6 observations per cluster period would be required to detect the same effect size for the cluster-level treatment, and 3 observations per cluster period required for detecting the same effect size for the individual-level treatment. To detect the same absolute value of interaction (when treatment conditions are coded as 0-1) would require 6 observations per cluster-period. These values are summarised in Table \ref{tab:1}.

If we assume a block exchangeable correlation structure instead of an exchangeable correlation structure, it is necessary to choose values for two parameters, the within-period ICC $\rhoC$ and the between period ICC $\rhoCT$. In order to make a choice that is consistent with a single ICC of $0.2$ from the exchangeable version, we will use Kasza \emph{et al.}~\cite{kasza2023does}, which provides a relationship between the ICC estimated assuming an exchangeable correlation structure with the pairs $(\rhoC,\rhoCT)$ that would be estimated from the same data if a block exchangeable correlation structure is assumed instead. Making an arbitrary but common assumption that $\frac{\rhoC}{\rhoCT}=0.8$, results in a within-period correlation of $\rhoCT=0.24$ and a between-period correlation of $\rhoC=0.192$. Then in the absence of an interaction between the two different interventions, the required sample size is given as follows: To detect an effect size of $0.35$ standard deviations for the cluster-level effect with a power of $80\%$ and a significance level of $0.05$, $5$ obervations in each cluster period are required. To detect the same effect size in the individual-level effect would require 2 observations in each cluster period. However, if an interaction between the two interventions was included in the model, then 7 observations per cluster period would be required to detect the same effect size for the cluster-level treatment, and 3 observations per cluster period required for detecting the same effect size for the individual-level treatment. To detect the same absolute value of interaction (when treatment conditions are coded as 0-1) would require 5 observations per cluster-period. In comparison to the exchangeable version, we require one more observation per cluster period for the cluster-level effect with or without interaction, while the rest of the required sample sizes remain the same. These values are also included in Table \ref{tab:1}. To get a sense of how required cluster period sizes differ when the targeted effect size is much smaller, Table \ref{tab:2} shows the required number of observations per cluster period when $\delta=0.2$ instead of $\delta=0.35$. It's clear that if the block exchangeable correlation structure is used then the number of observations required for the cluster-period level intervention is much greater than in the exchangeable case, and the cluster-period level design will be the determining factor for the required sample size if the magnitude of treatment effects for both interventions is similar.

\begin{figure}
\includegraphics[width=\textwidth]{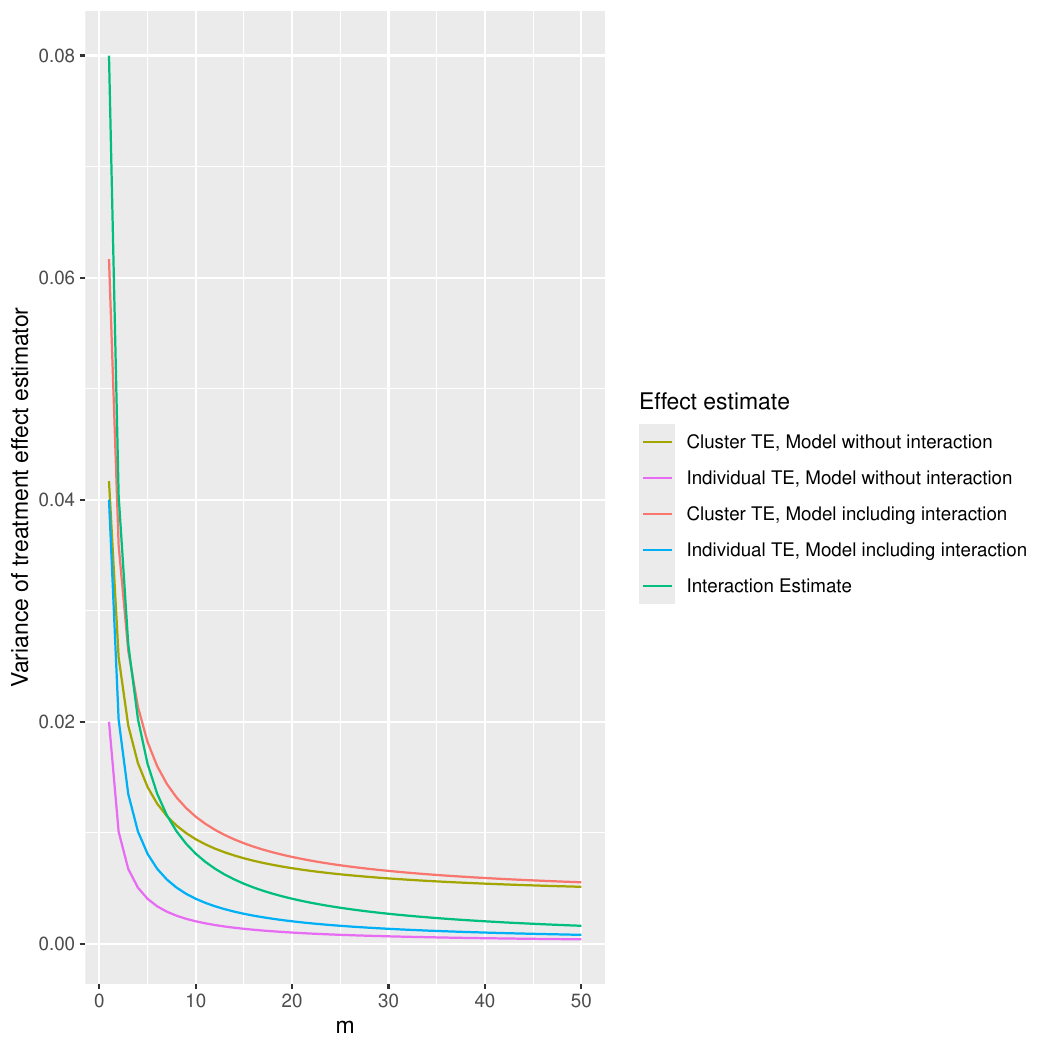}
\caption{Variance of treatment effect estimators vs number of observations per cluster period, for a standardised effect size of $\delta=0.2$, a block-exchangeable correlation structure with $\rhoCT=0.24$ and $\rhoC/\rhoCT=0.8$, and the other parameters as given in Section \ref{sec:example}.\label{fig:var}}
\end{figure}

\begin{figure}
\includegraphics[width=\textwidth]{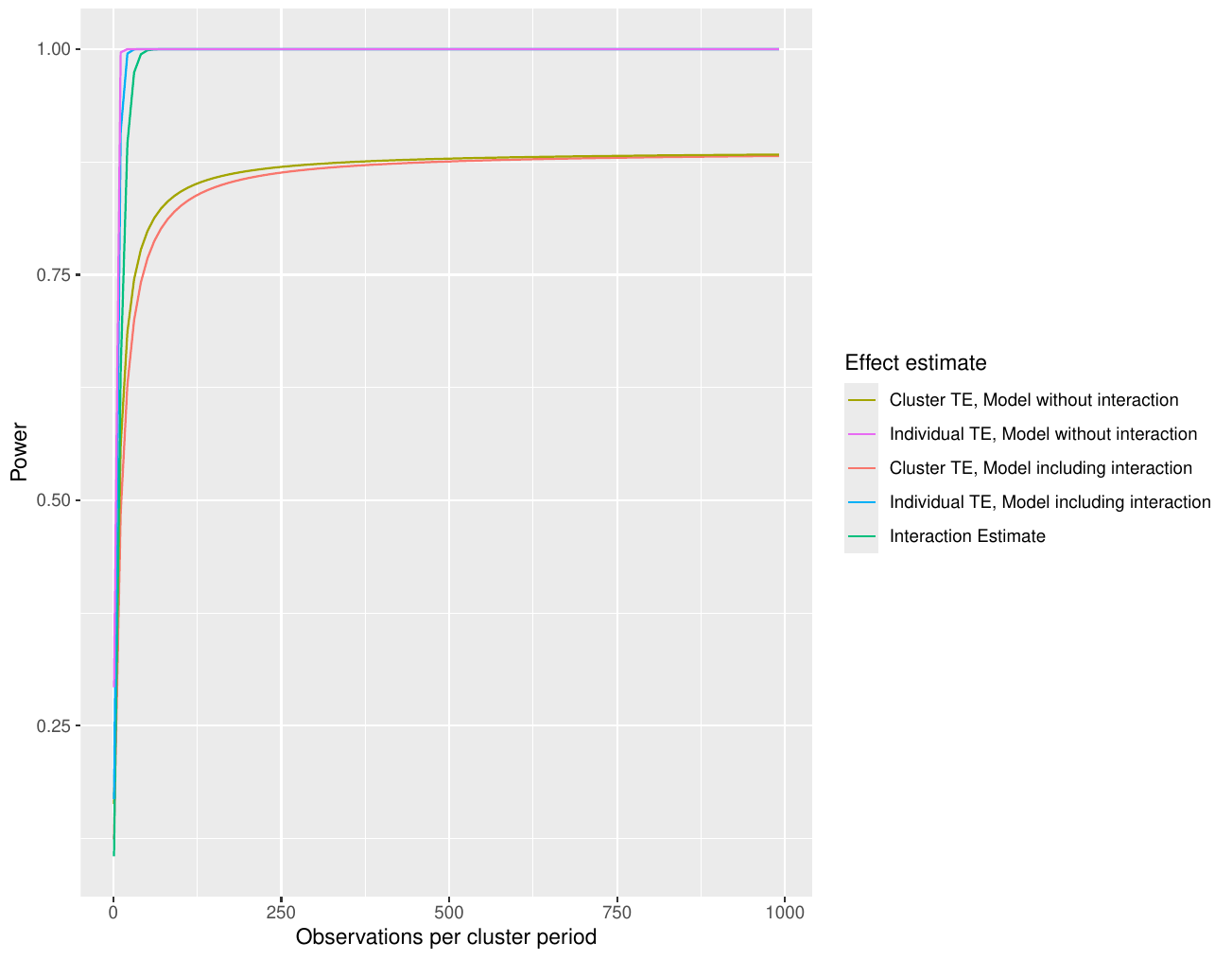}
\caption{Power vs number of observations per cluster period, for a standardised effect size of $\delta=0.2$, a block-exchangeable correlation structure with $\rhoCT=0.24$ and $\rhoC/\rhoCT=0.8$, and the other parameters as given in Section \ref{sec:example}.\label{fig:pow}.}
\end{figure}

\begin{figure}
\includegraphics[width=\textwidth]{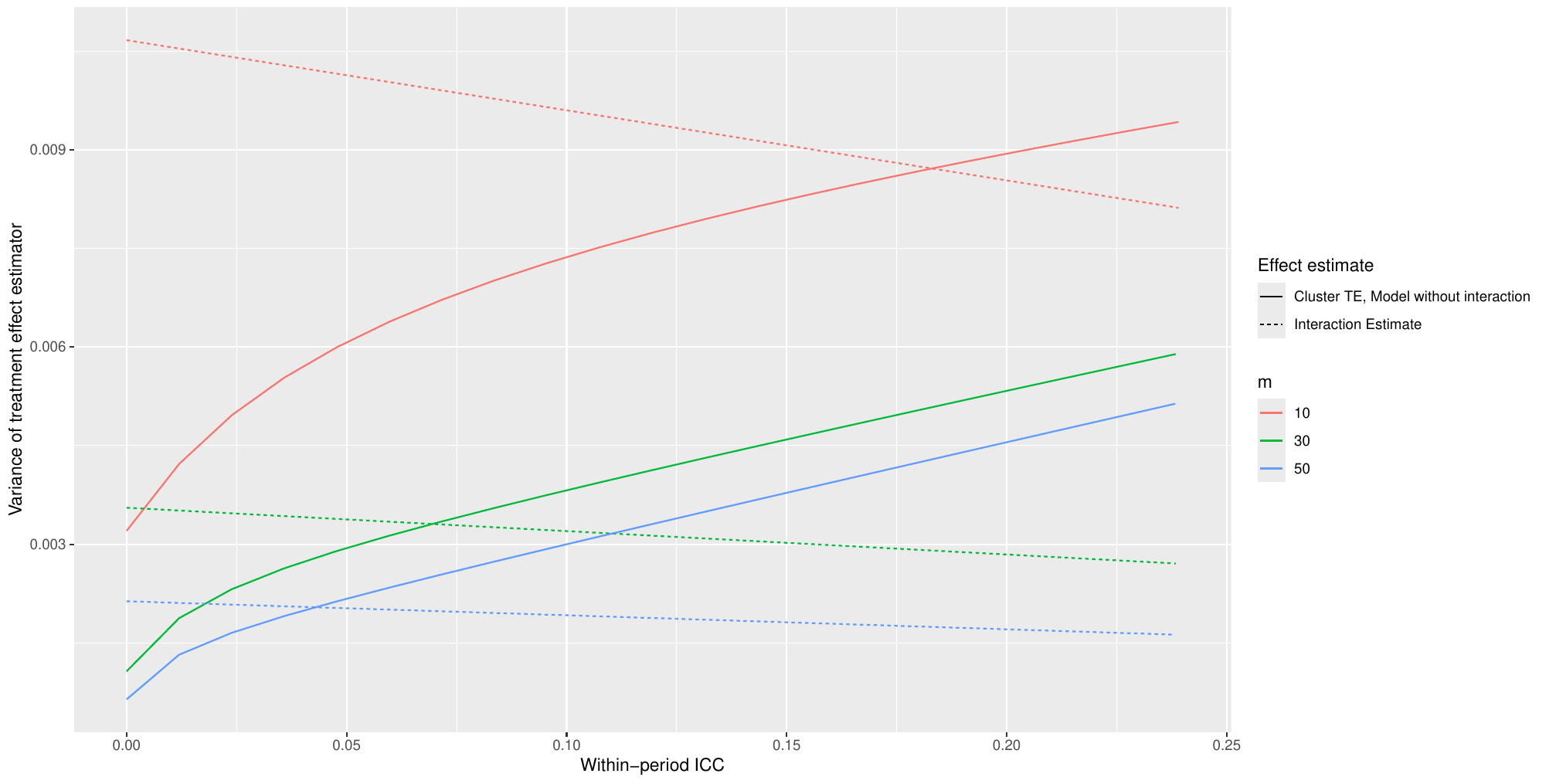}
\caption{Comparison of the variance of treatment effect estimators vs within-period ICC $\rhoCT$, for a standardised effect size of $\delta=0.2$, a block-exchangeable correlation structure $\rhoC/\rhoCT=0.8$, and the other parameters as given in Section \ref{sec:example}.\label{fig:varX}}
\end{figure}

Figures \ref{fig:var} and \ref{fig:pow} show how the power and variance of the estimates of the different treatment effects change as a function of the number of observations per cluster period, in the scenario given in this section. These show the typical behaviour characteristic of split-plot designs, where the estimate of the interaction term becomes more precise than that of the cluster treatment effect as the number of observations in each cluster-period increases. It is also true that as the within-period ICC increases (but $m$ is fixed), it has a similar effect, where the estimate of the interaction term becomes more precise than that of the cluster treatment effect, even compared to the cluster treatment effect in the model without an interaction term (Figure \ref{fig:varX}). This strength of split-plot designs often allows for interaction terms to be estimated much more precisely than they would be if both interventions were randomised at the cluster or cluster-period level.

\begin{table}[ht]
\begin{center}
\caption{Required number of observations per cluster period for the SharES trial under varying assumptions with a targeted effect size of $\delta=0.35$. \label{tab:1}}
\begin{tabular}{p{2.5cm} p{6cm} p{3cm} p{2.5cm} } 
\hline
\textbf{Interaction included}  &\textbf{Estimate} & \textbf{$m$, Exchangeable} & \textbf{$m$, Block exchangeable}\\
  \hline
Yes & Cluster-level intervention,      $\betaC$ & 6 & 7\\ 
  \hline
Yes & Individual-level intervention,  $\betaI$ & 3 & 3\\ 
  \hline
Yes & Interaction,                         $\betaCI$ & 6 & 5\\ 
  \hline
No &    Cluster-level intervention,   $\betaC$ & 4 & 5\\ 
      \hline
No &   Individual-level intervention, $\betaI$ & 2 & 2\\ 
  \hline
\end{tabular}
\end{center}
\end{table}

\begin{table}[ht]
\begin{center}
\caption{Required number of observations per cluster period for the SharES trial under varying assumptions with a targeted effect size of $\delta=0.2$. The exchangeable model has $\rhoCT=0.2$. The block exchangeable has $r=0.8$, $\rhoCT=0.24$. \label{tab:2}} 
\begin{tabular}{p{2.5cm} p{6cm} p{3cm} p{2.5cm} } 
\hline
\textbf{Interaction included}  &\textbf{Estimate} & \textbf{$m$, Exchangeable} & \textbf{$m$, Block exchangeable}\\
  \hline
Yes & Cluster-level intervention,      $\betaC$ & 21 & 71\\ 
  \hline
Yes & Individual-level intervention,  $\betaI$ & 11 & 11\\ 
  \hline
Yes & Interaction,                         $\betaCI$ & 21 & 21\\ 
  \hline
No &    Cluster-level intervention,   $\betaC$ & 21 & 61\\ 
      \hline
No &   Individual-level intervention, $\betaI$ & 11 & 11\\ 
  \hline
\end{tabular}
\end{center}
\end{table}

\section{Discussion and Conclusions}
This paper is the first to derive closed-form expressions for the variance of the treatment effect estimator and the required sample size for factorial trials that have an individually randomised treatment component, and a LCRT component (such as stepped wedge). Further, this expression is provided even in the case where cluster-period sizes are allowed to vary.

One limitation of this work is that it assumes either the nested exchangeable correlation structure or exchangeable correlation structure, and continuous outcomes. A further limitation is that it assumes block-randomisation of the individual-level treatment (\emph{i.e.} so that there are an equal number of intervention and control observations within each cluster-period), but this is common practice for modern individually randomised trials.

The results in this work are consistent with those for other types of split-plot design: both the variance of the individual-level treatment effect estimate and the estimate of the interaction term scale with the total number of observations, rather than the total number of clusters. This is very helpful when clusters are large but few in number, as is so often the case in longitudinal cluster randomised trials. One implication of this is that estimates of the interaction may be much more precise than those of the main cluster-period-level effect. Overall, in a trial with large numbers of individuals per cluster (as is often the case in longitudinal cluster randomised trials) it seems clear that the required sample size will be primarily dependent on the design for the LCRT-level treatment, rather than the individual-level treatment, assuming that effect sizes are similar for the two different interventions. 

Future work could focus on generalising to other types of outcome like binary or time-to-even outcomes; or allowing for different within-cluster correlation structures like discrete-time or continuous-time decay structures. These results might also be generalised to designs with a greater number of cluster- or individual-level factors, such as two cluster-level factors and one individual-level factor. Another generalisation of interest would be to designs with non-participation heterogeneity, where eligibility for treatments in one factor might depend on treatment allocation in another factor.

\bibliographystyle{plain}
\bibliography{bibliography}

\appendix
\section{Variance of treatment effect estimators -- Equal cluster period sizes and interaction term included}\label{sec:2}
We start by considering the case where there are equal cluster-period sizes, and an interaction term $\betaCI$ is included in the model. For the estimation of sample size, a vital component is the variance of the treatment effect estimator. In the case of multiple treatment efffects as we have here, the key matrix is the asymptotic (equivalent to feasible generalized least squares (FGLS) or generalized least squares (GLS) estimators conditional on the assumed variance/correlation parameters) covariance matrix of the effect estimators $\hbetaC,\hbetaI,\hbetaCI$. In order to find this matrix it is simpler to start by finding the covariance matrix of $\htbetaC,\hbetaI,\hbetaCI$ then derive the desired covariances with $\hbetaC$. We can also derive ``marginal'' treatment effects for any fixed level of prevalence of the other treatment. The desired matrix is given by the top left three rows and columns of
\begin{align}
(D^T \Sigma^{-1} D)^{-1} = (\sum_{i=1}^n D_i^T \Sigma_i^{-1} D_i)^{-1}
\end{align}
where $D$ is the design matrix for the (re-parametrised) model with the fixed effects $\beta$ in the order they are given in the model, $(\betaC,\betaI,\betaCI,\beta_{1},\ldots,\beta_{T})$, $D_i$ is the submatrix of $D$ with the rows corresponding to the $i$th cluster, $\Sigma$ is the covariance matrix of the vector $Y_{ijk}$ and $\Sigma_i$ is the within-cluster covariance matrix for $Y_{ijk}$ ordered lexicographically in terms of $(j,k)$, \emph{i.e.} individuals within periods. Note that $\Sigma_i$ is indexed by an $i$ for notational consistency, but it's the same for all clusters.

Let
\begin{align}
a_1 &= 1- \rhoCT\\
b_1 &= \rhoCT - \rhoC\\
c_1 &= \rhoC
\end{align}
so that
\begin{align}
\Sigma_i = \sigma^2(a_1 I_{Tm} + b_1 I_{T}\otimes J_m + c_1 J_{mT})
\end{align}
where $I_n$ and $J_n$ are the $n$ by $n$ identity matrix and matrix of all ones respectively, and $\otimes$ is the Kronecker product.

Then by the usual inversion formula:
\begin{align}
\Sigma_i^{-1} = \sigma^{-2}(a_2 I_{Tm} + b_2 I_{T}\otimes J_m + c_2 J_{mT})
\end{align}
where 
\begin{align}
a_2 &= \frac{1}{a_1}\\
b_2 &= \frac{-b_1}{a_1(a_1+b_1m)}\\
c_2 &= \frac{-c_1}{(a_1+b_1m)(a_1+b_1m+c_1mT)}
\end{align}

Then we break down $D_i^T \Sigma_i^{-1} D_i$ into $\sigma^{-2}a_2 D_i^T I_{Tm} D_i$ + $\sigma^{-2}b_2 D_i^T (I_{T}\otimes J_m) D_i$ + $\sigma^{-2}c_2 D_i^T J_{Tm}D_i$ and calculate each component separately (where rows/columns are ordered in the same order as the fixed effects are presented in the model above):
\begin{align}
D_i^T I_{Tm} D_i &= mT \begin{bmatrix}\pixi & 0 & 0 & X_{i1}/T &\ldots& X_{iT}/T\\
0 & \sigzz & \pixi \piz (1-\piz) & 0 &\cdots& 0\\
0 & \pixi \piz (1-\piz) & \pixi \piz (1-\piz) & 0 &\cdots& 0\\
X_{iT}/T & 0        & 0    & \frac{1}{T} & 0 & 0\\
\vdots & \vdots & \vdots & 0               &\ddots & 0\\
X_{iT}/T & 0       & 0     & 0 & 0 & \frac{1}{T}
\end{bmatrix}\label{eq1}\\
D_i^T (I_{T}\otimes J_m) D_i &= m^2T \begin{bmatrix}
\pixi & 0 & 0 & X_{i1}/T &\ldots& X_{iT}/T\\
0 & 0 & 0 & 0 &\cdots& 0\\
0 & 0 & 0 & 0 &\cdots& 0\\
X_{iT}/T & 0        & 0    & \frac{1}{T} & 0 & 0\\
\vdots & \vdots & \vdots & 0               &\ddots & 0\\
X_{iT}/T & 0       & 0     & 0 & 0 & \frac{1}{T}
\end{bmatrix}\label{eq2}\\
D_i^T J_{Tm} D_i &= m^2T^2 \begin{bmatrix}
\pixi & 0 & 0 & 1/T &\ldots& 1/T
\end{bmatrix}\begin{bmatrix}
\pixi & 0 & 0 & 1/T &\ldots& 1/T
\end{bmatrix}^T\label{eq3}
\end{align}
where $\sigzz = \piz(1-\piz)$. Then set 
\begin{align}
a_3 &= a_2mT\\
b_3 &= b_2m^2T\\
c_3 &= c_2m^2T^2
\end{align}
Then
\begin{align*}
&\sigma^2 D_i^T\Sigma_i^{-1}D_i = \\&\begin{bmatrix}
(a_3+b_3+c_3\pixi)\pixi &0 &0 & (a_3+b_3)\frac{X_{i1}}{T}+c_3\frac{\pixi}{T} &\ldots & (a_3+b_3)\frac{X_{iT}}{T}+c_3\frac{\pixi}{T}\\
0 & a_3\sigzz &a_3\pixi\sigzz & 0 &\ldots & 0\\
0 & a_3\pixi\sigzz &a_3\pixi\sigzz & 0 &\ldots & 0\\
(a_3+b_3)\frac{X_{i1}}{T}+c_3\frac{\pixi}{T} &0 &0 & \frac{1}{T}(a_3+b_3)+\frac{c_3}{T^2} &\cdots & \frac{c_3}{T^2}\\
\vdots & \vdots & \vdots & \vdots               &\ddots & \vdots\\
(a_3+b_3)\frac{X_{iT}}{T}+c_3 & \vdots & \vdots & \frac{c_3}{T^2}              &\cdots & \frac{1}{T}(a_3+b_3)+\frac{c_3}{T^2}
\end{bmatrix}
\end{align*}
where the bottom right $T$ by $T$ matrix corresponding to the time effects is $\frac{1}{T}(a_3+b_3)I_T+\frac{c_3}{T^2}J_T$.
Then we can reorder the rows/columns so that the first row/column is moved to after the second and third (\emph{i.e.} reorder the fixed effects in the order $\betaI,\betaCI,\tbetaC$):
\begin{align*}
&\sigma^2 D_i^T\Sigma_i^{-1}D_i = \\&\begin{bmatrix}
 a_3\sigzz &a_3\pixi\sigzz & 0 & 0 &\ldots & 0\\
 a_3\pixi\sigzz &a_3\pixi\sigzz & 0 & 0 &\ldots & 0\\
0 &0 &(a_3+b_3+c_3\pixi)\pixi & (a_3+b_3)\frac{X_{i1}}{T}+c_3\frac{\pixi}{T} &\ldots & (a_3+b_3)\frac{X_{iT}}{T}+c_3\frac{\pixi}{T}\\
0 &0 &(a_3+b_3)\frac{X_{i1}}{T}+c_3\frac{\pixi}{T} & \frac{1}{T}(a_3+b_3)+\frac{c_3}{T^2} &\cdots & \frac{c_3}{T^2}\\
\vdots & \vdots & \vdots & \vdots               &\ddots & \vdots\\
\vdots & \vdots & \vdots & \frac{c_3}{T^2}              &\cdots & \frac{1}{T}(a_3+b_3)+\frac{c_3}{T^2}
\end{bmatrix}
\end{align*}
which is a block-diagonal matrix (and thus we can invert the two blocks separately) and suggests labelling the top left hand block (the first two rows and columns) of $D_i^T\Sigma_iD_i$ as
\begin{align}
M_{1,i} = \frac{1}{\sigma^2}\begin{bmatrix}
a_3\sigzz &a_3\pixi\sigzz\\
 a_3\pixi\sigzz &a_3\pixi\sigzz
\end{bmatrix}
\end{align}
and the bottom right block as 
\begin{align}
M_{2,i} = \begin{bmatrix}
(a_3+b_3+c_3\pixi)\pixi & (a_3+b_3)\frac{X_{i1}}{T}+c_3\frac{\pixi}{T} &\ldots & (a_3+b_3)\frac{X_{iT}}{T}+c_3\frac{\pixi}{T}\\
(a_3+b_3)\frac{X_{i1}}{T}+c_3\frac{\pixi}{T} & \frac{1}{T}(a_3+b_3)+\frac{c_3}{T^2} &\cdots & \frac{c_3}{T^2}\\
 \vdots & \vdots               &\ddots & \vdots\\
 (a_3+b_3)\frac{X_{iT}}{T}+c_3\frac{\pixi}{T} & \frac{c_3}{T^2}              &\cdots & \frac{1}{T}(a_3+b_3)+\frac{c_3}{T^2}
\end{bmatrix}
\end{align}
Summing over $i$ we get:
\begin{align}
\sum_{i=1}^n M_{1,i} &= a_3 \sigma^{-2}\sigzz \sum_{i=1}^n\begin{bmatrix}
1&\pixi\\
 \pixi &\pixi
\end{bmatrix}\\
&= a_3 \sigma^{-2}\sigzz \begin{bmatrix}
n&n\pix\\
 n\pix &n\pix
\end{bmatrix}\\
&= a_3 n\sigma^{-2}\sigzz \begin{bmatrix}
1&\pix\\
 \pix &\pix
\end{bmatrix}\\
\therefore \left(\sum_{i=1}^n M_{1,i}\right)^{-1} &=  \frac{1}{a_3\sigzz} \frac{\sigma^2}{n}
\begin{bmatrix}
\pix&-\pix\\
 -\pix &1
 \end{bmatrix}
\end{align}
and
\begin{align}
\sum_{i=1}^n M_{2,i}&= \begin{bmatrix}
(a_3+b_3)\sum_i \pixi+c_3\sum_i\pixi^2 & \sum_i ((a_3+b_3)\frac{X_{i1}}{T}+c_3\frac{\pixi}{T}) &\ldots & \sum_i((a_3+b_3)\frac{X_{iT}}{T}+c_3\frac{\pixi}{T})\\
\sum_i ((a_3+b_3)\frac{X_{i1}}{T}+c_3\frac{\pixi}{T}) & n(\frac{1}{T}(a_3+b_3)+\frac{c_3}{T^2}) &\cdots & n\frac{c_3}{T^2}\\
 \vdots & \vdots               &\ddots & \vdots\\
 \vdots & n\frac{c_3}{T^2}              &\cdots & n(\frac{1}{T}(a_3+b_3)+\frac{c_3}{T^2})
\end{bmatrix}\\
&= n\begin{bmatrix}
(a_3+b_3)\pix+c_3\pixx & (a_3+b_3)\frac{\pi_{X,\cdot 1}}{T}+c_3\frac{\pix}{T} &\ldots & (a_3+b_3)\frac{\pi_{X,\cdot T}}{T}+c_3\frac{\pix}{T}\\
(a_3+b_3)\frac{\pi_{X,\cdot 1}}{T}+c_3\frac{\pix}{T} & \frac{1}{T}(a_3+b_3)+\frac{c_3}{T^2} &\cdots & \frac{c_3}{T^2}\\
 \vdots & \vdots               &\ddots & \vdots\\
 \vdots & \frac{c_3}{T^2}              &\cdots & \frac{1}{T}(a_3+b_3)+\frac{c_3}{T^2}
\end{bmatrix}
\end{align}
where $\pixx=\frac{1}{n}\sum_{i=1}^n\pixi^2$ and $\pi_{X,\cdot j} = \frac{1}{n}\sum_{i=1}^nX_{ij}$. But $\sum_{i=1}^n M_{2,i}$ is just the same form as the equivalent one for a standard longitudinal cluster randomised trial with the same design (but without individual-level treatments). This is clear because it is just the matrix we would have obtained if we restricted the model to not include $Z_{i,j,k}$ or $X_{i,j}Z_{i,j,k}$ terms but kept everything else the same. Thus the estimator $\htbetaC$ (for the marginal treatment effect for the cluster-period-level treatment) has the same variance as that for $\hat\theta$ in a standard LCRT of the same design, which we will denote $\VL$. Moreover, it is uncorrelated with the estimates $\hbetaI$ and $\hbetaCI$. Formulas for $\VL$ are well known in a wide variety of cases: for example see~\cite{Hooper2016}. One explicit expression for $\VL$ is given in Section \ref{sec:sample_size}.

Thus
\begin{align}
\cov(\hat{\boldsymbol{\beta}}|D,\sigma^2,\rhoC,\rhoCT) = \sigma^2 %
\left[\begin{array}{cccc}
\multicolumn{2}{c}{\multirow{2}{*}{\large $\left(\sum_{i=1}^n M_{1,i}\right)^{-1}$}} & 0 & \cdots \\
 & & 0 & \cdots\\
0 & 0 & \multicolumn{2}{c}{\multirow{2}{*}{\large $\left(\sum_{i=1}^n M_{2,i}\right)^{-1}$}}\\
\vdots & \vdots &  & \\
\end{array}\right]
\end{align}

Pulling out the entries corresponding to the estimators we get
\begin{align}
\var(\htbetaC) &= \VL\\
\var(\hbetaI) &= \frac{\pix}{a_3\sigzz\pix(1-\pix)}\cdot\frac{\sigma^2}{n}= \frac{(1-\rhoCT)}{mT\sigzz(1-\pix)}\cdot\frac{\sigma^2}{n}\\
\var(\hbetaCI) &= \frac{1}{a_3\sigzz\pix(1-\pix)}\cdot\frac{\sigma^2}{n} = \frac{(1-\rhoCT)}{mT\sigzz\pix(1-\pix)}\cdot\frac{\sigma^2}{n}\\
\var(\hbetaC) &= \var(\htbetaC-\hbetaCI\piz)\notag\\
&=\var(\htbetaC)+\piz^2\var(\hbetaCI)-2\cov(\htbetaC,\hbetaCI\piz)\notag\\
&=\VL + \frac{\piz^2(1-\rhoCT)}{mT\sigzz\pix(1-\pix)}\cdot\frac{\sigma^2}{n}.\label{eq:beta2_int}
\end{align}

Note that $\var(\hbetaI) = \pix\var(\hbetaCI)$. Also, $\rhoC$ is not involved in the expressions for $\var(\hbetaI)$ and $\var(\hbetaCI)$.

\section{Treatment effect estimator variances (without interaction) -- equal cluster period sizes}\label{sec:no_int}
We can now use the derivation above to also allow us to calculate the variances of the treatment effect estimates when there is assumed to be no interaction between the interventions. For no interaction (no $\betaCI$ term in the model) then we lose a row and column of the inverse covariance matrix (of the estimators), and also $\tbetaC=\betaC$ now. So
\begin{align}
\sum_{i=1}^nM_{1,i} = a_3 \piz (1-\piz)\sigma^{-2}n\\
(\sum_{i=1}^nM_{1,i})^{-1} = \frac{\sigma^2}{a_3 \piz (1-\piz) n}
\end{align}
and thus
\begin{align}
\var(\hbetaI) &= \frac{\sigma^2}{a_3 \piz (1-\piz) n} = \frac{1-\rhoCT}{mT \sigma_Z^2}\cdot\frac{\sigma^2}{n} \\
\var(\hbetaC) &= \VL
\end{align}
We can compare the expressions for the variance of the individual TE with those in the model including interaction, and we see that 
\begin{align}
\var(\hat\beta_{\mathrm{I},\mathrm{interaction}}) = \frac{1}{1-\pix}\var(\hat\beta_{\mathrm{I},\mathrm{nointeraction}}).
\end{align}
Since $0<\pix<1$, $\var(\hat\beta_{\mathrm{I},\mathrm{interaction}})>\var(\hat\beta_{\mathrm{I},\mathrm{nointeraction}})$, unsurprisingly.

\section{Comparing all four treatment conditions}
It might be the case that we are interested in more complicated research questions that those assumed above. For instance, if we are interested in the best of the four possible treatment conditions, then we could be interested in the three constrasts $\betaC,\betaI,\betaC+\betaI+\betaCI$ against $X_{ij}=0,Z_{ijk}=0$. We know the covariance matrix for the initial estimates is:
\begin{align}
\cov(\hbetaI,\hbetaCI,\htbetaC) &= 
\begin{bmatrix}
K\pix  & -K\pix & 0\\
-K\pix & K      & 0\\
0     & 0     & \VL
 \end{bmatrix} 
\end{align}
where $K = \frac{1}{a_3\sigzz} \frac{\sigma^2}{n}$.
Each contrast we are interested in is a linear combination of those estimates:
\begin{align}
[\hbetaI,\hbetaC,\hbetaC+\hbetaI+\hbetaCI]^T = \begin{bmatrix}
1 & 0 & 0\\
0 & -\piz & 1\\
1 & 1-\piz & 1
\end{bmatrix}\begin{bmatrix}
\hbetaI\\\hbetaCI\\\htbetaC
\end{bmatrix}
\end{align}
Therefore the covariance matrix for the three contrasts is (when an interaction term is included)
\begin{align*}
\cov(\hbetaI,\hbetaC,\hbetaC+\hbetaI+\hbetaCI) &= \begin{bmatrix}
1 & 0 & 0\\
0 & -\piz & 1\\
1 & 1-\piz & 1
\end{bmatrix}
 \begin{bmatrix}
K\pix  & -K\pix & 0\\
-K\pix & K      & 0\\
0     & 0     & \VL
 \end{bmatrix} 
\begin{bmatrix}
1 & 0 & 0\\
0 & -\piz & 1\\
1 & 1-\piz & 1
\end{bmatrix}^T\\
&= \begin{bmatrix} 
K\pix & K\pix\piz & K\pix\piz\\
K\pix\piz & K\pix\piz+V & K\pix(2\piz-1)+V\\
K\pix\piz & K\pix(2\piz-1)+V & K(\pix(2\piz-1)+(1-\piz)^2)+V
\end{bmatrix}
\end{align*}
which is the important matrix for the joint distribution of the three different treatment effect estimates. This can then be used to derive powers or sample sizes for more complicated tests.

\section{Variable cluster-period sizes}\label{sec:var_size}
In the previous derivations we assumed that there are the same number of observations in each cluster-period cell. For clinical trials for example this is rarely the case. So we now consider the case where there are a random number $m_{i,j}$ of observations in each cluster-period $(i,j)$. Even when there are a random number of observations in each cluster period, it is common for trials with individual randomisation to do block-randomisation, thus guaranteeing that the desired proportion of individuals are allocated to each treatment condition within each randomisation block (and hence overall also). So we assume that $(1-\pi_Z)m_{i,j}$ of them correspond to $Z=0$ and $\pi_Zm_{i,j}$ of them have $Z=1$ in each cluster period. This means that while we allow the number of participants in each cluster-period to vary, the proportion of them allocated to each individual-level treatment is fixed.

Note that each observation with the same cluster $i$, period $j$ and value of $Z_{i,j,k}$ is exchangeable under our model. The GLS estimator is also a linear estimator. Thus, if we let $\ell=Z_{i,j,k}$ we know that the estimator will have the same coefficient for each observation with the same value of $(i,j,\ell)$, and so we need only consider means of observations within each group of $(i,j,\ell)$, \emph{i.e.} each cluster-period-$Z$ group. That is, let
\begin{align*}
\bar{Y}_{i,j,\cdot,0} &= \frac{1}{(1-\pi_Z)m_{i,j}} \sum_{k|Z_{i,j,k}=0} Y_{i,j,k}\\
\bar{Y}_{i,j,\cdot,1} &= \frac{1}{\pi_Zm_{i,j}} \sum_{k|Z_{i,j,k}=1} Y_{i,j,k}
\end{align*}
 
 Our general approach will be to split the covariance matrix of these means into the sum of a block diagonal (and hence invertible) matrix, and a constant matrix expressed as an outer product of two vectors; then use the Sherman Morrison 
  formula to invert that matrix sum into one of the same form; and then evaluate $D_i^T\Sigma_i^{-1}D_i$ by evaluating the components corresponding to the block diagonal and constant matrix separately. Once that is done we can find $(\sum_{i}D_i^T\Sigma_i^{-1}D_i)^{-1}$ to get the variances for the treatment effects, which is relatively simple in this case because of the structure of $\sum_{i}D_i^T\Sigma_i^{-1}D_i$.
 
 So we start by writing out the covariance matrix for those $(i,j,\ell)$-means for cluster $i$, \emph{i.e.}
 \begin{align*}
&\Sigma_i = \cov((\bar{Y}_{i,j,\cdot,\ell})_{j=1,\ldots,T;\ell=0,1}) =\\
 &\begin{bmatrix}\sigma_C^2+\sigma_{CT}^2+\frac{\sigma_{\varepsilon}^2}{(1-\pi_Z)m_{i,1}} & \sigma_C^2+\sigma_{CT}^2 & \sigma_C^2 & \sigma_C^2 & \sigma_C^2 & \ldots\\
\sigma_C^2+\sigma_{CT}^2 & \sigma_C^2+\sigma_{CT}^2 +\frac{\sigma_{\varepsilon}^2}{\pi_Zm_{i,1}}& \sigma_C^2 & \sigma_C^2 & \sigma_C^2 & \ldots\\
\sigma_C^2 & \sigma_C^2 & \sigma_C^2+\sigma_{CT}^2+\frac{\sigma_{\varepsilon}^2}{(1-\pi_Z)m_{i,2}} & \sigma_C^2+\sigma_{CT}^2 & \sigma_C^2 & \ldots\\
\sigma_C^2 & \sigma_C^2 & \sigma_C^2+\sigma_{CT}^2 & \sigma_C^2+\sigma_{CT}^2 +\frac{\sigma_{\varepsilon}^2}{\pi_Zm_{i,2}}& \sigma_C^2 & \ldots \\
\sigma_C^2 & \sigma_C^2 & \sigma_C^2 & \sigma_C^2 & \ddots
\end{bmatrix}
\end{align*}
 
 Let 
\begin{align*}
&A_i =\\
 &\begin{bmatrix}\sigma_{CT}^2+\frac{\sigma_{\varepsilon}^2}{(1-\pi_Z)m_{i,1}} & \sigma_{CT}^2 & 0 & 0 & 0 & \ldots\\
\sigma_{CT}^2 &\sigma_{CT}^2 +\frac{\sigma_{\varepsilon}^2}{\pi_Zm_{i,1}}& 0 & 0 & 0 & \ldots\\
0 & 0 &\sigma_{CT}^2+\frac{\sigma_{\varepsilon}^2}{(1-\pi_Z)m_{i,2}} & \sigma_{CT}^2 & 0 & \ldots\\
0 & 0 & \sigma_{CT}^2 & \sigma_{CT}^2 +\frac{\sigma_{\varepsilon}^2}{\pi_Zm_{i,2}}& 0 & \ldots \\
0 & 0 & 0 & 0 & \ddots
\end{bmatrix}
\end{align*}
so that $\Sigma_i = A_i + \sigma^2_C \mathbf{1}_{2T}\mathbf{1}_{2T}^T$, where $\mathbf{1}_{2T}$ is the column vector of length $2T$ consisting entirely of ones. 
Now let $x_{i,j,1}=\frac{\sigma_{\varepsilon}^2}{(1-\pi_Z)m_{i,j}}$, $x_{i,j,2}=\frac{\sigma_{\varepsilon}^2}{\pi_Zm_{i,j}}$ and 
\begin{align}
A_{i,j}&=\begin{bmatrix}\sigma_{CT}^2+\frac{\sigma_{\varepsilon}^2}{(1-\pi_Z)m_{i,j}} & \sigma_{CT}^2\\
\sigma_{CT}^2 & \sigma_{CT}^2 +\frac{\sigma_{\varepsilon}^2}{\pi_Zm_{i,j}}
\end{bmatrix}\\ &= \begin{bmatrix}\bb+x_{i,j,1} & \bb \\
\bb & \bb + x_{i,j,2}
\end{bmatrix}
\end{align}
so that 
$A_i$ is the block diagonal matrix with blocks $(A_{i,j})_{j=1,\ldots,T}$, which we denote $A_i=\mathrm{diag}_{j=1,\ldots,T}(A_{i,j})$.
Thus, 
\begin{align}
A_i^{-1} &= \mathrm{diag}_{j=1,\ldots,T}(A_{i,j}^{-1})\\
&= \mathrm{diag}_{j=1,\ldots,T}\left(\frac{1}{|A_{i,j}|}\begin{bmatrix}
\bb+x_{i,j,2} & -\bb\\
-\bb & \bb + x_{i,j,1}
\end{bmatrix}\right)
\end{align}
where $|A_{i,j}|=(\bb+x_{i,j,1})(\bb+x_{i,j,2})-(\bb)^2=\bb(x_{i,j,1}+x_{i,j,2})+x_{i,j,1}x_{i,j,2} $

As per usual, we wish to find \begin{align}
\var(\hat{\boldsymbol{\beta}}) = (D^T\Sigma^{-1} D)^{-1} = (\sum_i D_i^T\Sigma_i^{-1} D_i)^{-1}
\end{align}
 where $D_i$ is the design matrix with only the rows corresponding to the observations (in this case the observations are the means $\bar{Y}_{i,j,\cdot,\ell}$) in the $i$th cluster. We start by finding $\Sigma_i^{-1} = (A_i + \sigma^2_C \mathbf{1}_{2T}\mathbf{1}_{2T}^T)^{-1}$. Applying the Sherman-Morrison formula to our definition of $\Sigma_i$ we have 
\begin{align}
\Sigma_i^{-1} &= (A_i + \sigma^2_C \mathbf{1}_{2T}\mathbf{1}_{2T}^T)^{-1}\\
&= A_i^{-1}- \frac{A_i^{-1}\sigma^2_C \mathbf{1}_{2T}\mathbf{1}_{2T}^T A_i^{-1}}{1+\sigma^2_C \mathbf{1}_{2T}^TA_i^{-1}\mathbf{1}_{2T}}\\
&= A_i^{-1}- \frac{\sigma^2_C A_i^{-1} \mathbf{1}_{2T}\mathbf{1}_{2T}^T A_i^{-1}}{1+\sigma^2_C \sum_{j,\ell}\frac{x_{i,j,\ell}}{|A_{i,j}|}}\\
&= A_i^{-1}- \frac{\sigma^2_C (x_{i,1,2},x_{i,1,1},x_{i,2,2},\ldots)^T(x_{i,1,2},x_{i,1,1},x_{i,2,2},\ldots)}{1+\sigma^2_C \sum_{j,\ell}\frac{x_{i,j,\ell}}{|A_{i,j}|}}\\
&= A_i^{-1}- \frac{\sigma^2_C \xx_i \xx_i^T}{1+\sigma^2_C \sum_{j,\ell}\frac{x_{i,j,\ell}}{|A_{i,j}|}}\label{eq:sigi_inv}
\end{align}
where $\xx_i=(\frac{x_{i,1,2}}{|A_{i,1}|},\frac{x_{i,1,1}}{|A_{i,1}|},\frac{x_{i,2,2}}{|A_{i,2}|},\ldots)^T$.

Now in order to find $(D^T\Sigma^{-1} D)^{-1} = (\sum_i D_i^T\Sigma_i^{-1} D_i)^{-1}$ we split (\ref{eq:sigi_inv}) into two parts and evaluate $\dd_{c_1}^TA_i^{-1}\dd_{c_2}$ and $\dd_{c_1}^T\xx_i\xx_i^T\dd_{c_2} = (\dd_{c_1}^T\xx_i)(\dd_{c_2}^T\xx_i)^T$ for each pair of columns $\dd_{c_1},\dd_{c_2}$ of $D_i$; with $c_1,c_2=1,\ldots,T+3$. Since each column of $D$ and $D_i$ corresponds to a fixed effect, we will label them correspondingly, so that $D=[\dd_Z,\dd_{XZ},\dd_X,\dd_{j=1},\dd_{j=2},\ldots,\dd_{j=T}]$ retaining the ordering of columns used at the end of Appendix \ref{sec:2}, and drop the index $i$ for clarity (leaving it implied by context). Thus,
\begin{align}
\dd_Z &= (-\pi_Z,1-\pi_Z,-\pi_Z,1-\pi_Z,\ldots)^T\\
\dd_{XZ} &= (-\pi_ZX_{i1},(1-\pi_Z)X_{i1},-\pi_ZX_{i2},(1-\pi_Z)X_{i2},\ldots)^T\\
\dd_X &= (X_{i1},X_{i1},X_{i2},X_{i2},X_{i3},\ldots)^T\\
\dd_{j=1} &= (1,1,0,0,0,\ldots)^T\\
&\vdots\\
\dd_{j=T} &= (0,0,\ldots,0,1,1)^T
\end{align}
and using $\pi_Z^2x_{i,j,2}+(1-\pi_Z)^2x_{i,j,1} = \pi_Z^2\frac{\sve}{\pi_Zm_{i,j}}+(1-\pi_Z)^2\frac{\sve}{(1-\pi_Z)m_{i,j}} = \frac{\sve}{m_{i,j}}$ and also $x_{i,j,1}+x_{i,j,2} = \frac{\sve}{m_{i,j}}\left(\frac{1}{\pi_Z}+\frac{1}{1-\pi_Z}\right) = \frac{\sve}{m_{i,j}}\left(\frac{1}{\pi_Z(1-\pi_Z)}\right) = \frac{\sve}{m_{i,j}\sigma_Z^2}$ we have that
\begin{align}
\dd_Z^T A_i^{-1}\dd_Z &= \sum_{j=1}^T 
\frac{1}{|A_{i,j}|}
\begin{bmatrix}
-\pi_Z & 1-\pi_Z 
\end{bmatrix}
\begin{bmatrix}
x_{i,j,2}+\sigma_{CT}^2 & -\sigma_{CT}^2\\ 
-\sigma_{CT}^2 & x_{i,j,1}+\sigma_{CT}^2
\end{bmatrix}
\begin{bmatrix}
-\pi_Z \\ 1-\pi_Z 
\end{bmatrix}\\
&= \sum_{j=1}^T \frac{1}{|A_{i,j}|} (\pi_Z^2x_{i,j,2}+(1-\pi_Z)^2x_{i,j,1}+\sigma_{CT}^2(\pi_Z^2+2(-1)(-\pi_Z)(1-\pi_Z)+(1-\pi_Z)^2))\\
&= \sum_{j=1}^T \frac{1}{|A_{i,j}|} (\pi_Z^2x_{i,j,2}+(1-\pi_Z)^2x_{i,j,1}+\sigma_{CT}^2)\\
&=\sum_{j=1}^T \frac{1}{|A_{i,j}|} \left(\frac{\sve}{m_{i,j}}+\sigma_{CT}^2\right)\\
&=\sum_{j=1}^T \frac{\sigma^2_{i,j}}{|A_{i,j}|}
\end{align}
where $\sigma^2_{i,j}=\sigma_{CT}^2+\frac{\sve}{m_{i,j}}$. Continuing, we have
\begin{align}
\dd_{X}^T A_i^{-1}\dd_Z &= \sum_{j=1}^T 
\frac{1}{|A_{i,j}|}
\begin{bmatrix}
X_{i,j}& X_{i,j} 
\end{bmatrix}
\begin{bmatrix}
x_{i,j,2}+\bb & -\sigma_{CT}^2\\ 
-\sigma_{CT}^2 & x_{i,j,1}+\sigma_{CT}^2
\end{bmatrix}
\begin{bmatrix}
-\pi_Z \\ 1-\pi_Z 
\end{bmatrix}\\
&= \sum_{j=1}^T \frac{X_{i,j}}{|A_{i,j}|}(-\pi_Zx_{i,j,2}+(1-\pi_Z)x_{i,j,1})\\
&= \sum_{j=1}^T \frac{X_{i,j}}{|A_{i,j}|}(-\pi_Z \left(\frac{1-\pi_Z}{\pi_Z}\right)x_{i,j,1}+(1-\pi_Z)x_{i,j,1})\\
&= 0\\
\dd_{XZ}^T A_i^{-1}\dd_{Z} &= \sum_{j=1}^T 
\frac{1}{|A_{i,j}|}
X_{i,j}\begin{bmatrix}
-\pi_Z & 1-\pi_Z 
\end{bmatrix}
\begin{bmatrix}
x_{i,j,2}+\sigma_{CT}^2 & -\sigma_{CT}^2\\ 
-\sigma_{CT}^2 & x_{i,j,1}+\sigma_{CT}^2
\end{bmatrix}
\begin{bmatrix}
-\pi_Z \\ 1-\pi_Z 
\end{bmatrix}\\
&= \sum_{j=1}^T \frac{X_{i,j}}{|A_{i,j}|} (\pi_Z^2x_{i,j,2}+(1-\pi_Z)^2x_{i,j,1}+\sigma_{CT}^2(\pi_Z^2+2\pi_Z(1-\pi_Z)+(1-\pi_Z)^2))\\
&= \sum_{j=1}^T \frac{X_{i,j}\sigma^2_{i,j}}{|A_{i,j}|} \\
\dd_{j=t}^T A_i^{-1}\dd_Z &= \sum_{j=1}^T 
\frac{1_{\{j=t\}}}{|A_{i,j}|}
\begin{bmatrix}
1&1
\end{bmatrix}
\begin{bmatrix}
x_{i,j,2}+\bb & -\sigma_{CT}^2\\ 
-\sigma_{CT}^2 & x_{i,j,1}+\sigma_{CT}^2
\end{bmatrix}
\begin{bmatrix}
-\pi_Z \\ 1-\pi_Z 
\end{bmatrix}\\
&= \frac{1}{|A_{i,t}|}(-\pi_Zx_{i,t,2}+(1-\pi_Z)x_{i,t,1})\\
&= \frac{1}{|A_{i,t}|}(-\pi_Z \left(\frac{1-\pi_Z}{\pi_Z}\right)x_{i,t,1}+(1-\pi_Z)x_{i,t,1})\\
&= 0\end{align}
\begin{align}
\dd_{X}^T A_i^{-1}\dd_X &= \sum_{j=1}^T 
\frac{1}{|A_{i,j}|}
\begin{bmatrix}
X_{i,j}& X_{i,j} 
\end{bmatrix}
\begin{bmatrix}
x_{i,j,2}+\sigma_{CT}^2 & -\sigma_{CT}^2\\ 
-\sigma_{CT}^2 & x_{i,j,1}+\sigma_{CT}^2
\end{bmatrix}
\begin{bmatrix}
X_{i,j}\\ X_{i,j} 
\end{bmatrix}\\
&=  \sum_{j=1}^T \frac{X_{i,j}}{|A_{i,j}|}(x_{i,j,2}+x_{i,j,1})\\
&=\sum_{j=1}^T \frac{X_{i,j}\sve}{|A_{i,j}|m_{i,j}\sigma_Z^2}\\
\dd_{XZ}^T A_i^{-1}\dd_{X} &= \sum_{j=1}^T 
\frac{1}{|A_{i,j}|}
\begin{bmatrix}
X_{i,j}& X_{i,j} 
\end{bmatrix}
\begin{bmatrix}
x_{i,j,2}+\sigma_{CT}^2 & -\sigma_{CT}^2\\ 
-\sigma_{CT}^2 & x_{i,j,1}+\sigma_{CT}^2
\end{bmatrix}
\begin{bmatrix}
X_{i,j}(-\pi_Z)\\ X_{i,j}(1-\pi_Z)
\end{bmatrix}\\
&= \sum_{j=1}^T \frac{X_{i,j}}{|A_{i,j}|}(-\pi_Zx_{i,j,2}+(1-\pi_Z)x_{i,j,1})\qquad\mbox{(same as $\dd_{X}^T A_i^{-1}\dd_Z$ since $X_{i,j}^2=X_{i,j}$) }\\
&= 0\\
\dd_{j=t}^T A_i^{-1}\dd_X &= \sum_{j=1}^T 
\frac{1_{\{j=t\}}}{|A_{i,j}|}
\begin{bmatrix}
1&1
\end{bmatrix}
\begin{bmatrix}
x_{i,j,2}+\sigma_{CT}^2 & -\sigma_{CT}^2\\ 
-\sigma_{CT}^2 & x_{i,j,1}+\sigma_{CT}^2
\end{bmatrix}
\begin{bmatrix}
X_{i,j}\\ X_{i,j} 
\end{bmatrix}\\
&= \frac{X_{i,t}}{|A_{i,t}|}(x_{i,t,2}+x_{i,t,1})\\
&= \frac{X_{i,t}\sve}{|A_{i,t}|m_{i,t}\sigma_Z^2}\\
\dd_{XZ}^T A_i^{-1}\dd_{XZ} &= \sum_{j=1}^T 
\frac{1}{|A_{i,j}|}
X_{i,j}\begin{bmatrix}
-\pi_Z & 1-\pi_Z 
\end{bmatrix}
\begin{bmatrix}
x_{i,j,2}+\sigma_{CT}^2 & -\sigma_{CT}^2\\ 
-\sigma_{CT}^2 & x_{i,j,1}+\sigma_{CT}^2
\end{bmatrix}
\begin{bmatrix}
-\pi_Z \\ 1-\pi_Z 
\end{bmatrix}\\
&= \sum_{j=1}^T \frac{X_{i,j}}{|A_{i,j}|} (\pi_Z^2x_{i,j,2}+(1-\pi_Z)^2x_{i,j,1}+\sigma_{CT}^2(\pi_Z^2+2(-1)(-\pi_Z)(1-\pi_Z)+(1-\pi_Z)^2))\\
&= \sum_{j=1}^T \frac{X_{i,j}\sigma^2_{i,j}}{|A_{i,j}|}\\
\dd_{j=t}^T A_i^{-1}\dd_{XZ} &= \sum_{j=1}^T 
\frac{1_{\{j=t\}}}{|A_{i,j}|}X_{i,j}
\begin{bmatrix}
1&1
\end{bmatrix}
\begin{bmatrix}
x_{i,j,2}+\sigma_{CT}^2 & -\sigma_{CT}^2\\ 
-\sigma_{CT}^2 & x_{i,j,1}+\sigma_{CT}^2
\end{bmatrix}
\begin{bmatrix}
-\pi_Z \\ 1-\pi_Z 
\end{bmatrix}\\
&= \frac{X_{i,t}}{|A_{i,t}|}(-\pi_Zx_{i,t,2}+(1-\pi_Z)x_{i,t,1})\\
&= \frac{X_{i,t}}{|A_{i,t}|}(-\pi_Z \left(\frac{1-\pi_Z}{\pi_Z}\right)x_{i,t,1}+(1-\pi_Z)x_{i,t,1})\\
&= 0\end{align}
We also need to evaluate $\dd_{c}\xx$ for each column $\dd_c$ of $D_i$.
\begin{align}
\dd_Z^T\xx &= \sum_j \frac{1}{|A_{i,j}|}\left((-\pi_Z)x_{i,j,2}+(1-\pi_Z)x_{i,j,1}\right)\\
&=\sum_j \frac{1}{|A_{i,j}|}\left(-\frac{\sve}{m_{i,j}}+\frac{\sve}{m_{i,j}}\right)\\
&=0
\\
\dd_{XZ}^T\xx &= \sum_j \frac{X_{i,j}}{|A_{i,j}|}\left((-\pi_Z)x_{i,j,2}+(1-\pi_Z)x_{i,j,1}\right)\\
&=\sum_j \frac{X_{i,j}}{|A_{i,j}|}\left(-\frac{\sve}{m_{i,j}}+\frac{\sve}{m_{i,j}}\right)\\
&=0
\end{align}

Note that this means that the first two columns and first two rows of $(D^Tq)(D^Tq)^T$ will be zero. Combined with the other zeros of $D^TA_i^{-1}D$ from above, we have the same structure of $D^T\Sigma_i^{-1}D$ as in Appendix \ref{sec:2} where it is block diagonal with a 2 by 2 upper left block $M_{1,i}$ and a $T+1$ by $T+1$ lower right block, with the remainder of the matrix equal to 0, \emph{i.e.}
\begin{align}
D^T\Sigma_i D = %
\left[\begin{array}{cccc}
\multicolumn{2}{c}{\multirow{2}{*}{\large $M_{1,i}$}} & 0 & \cdots \\
 & & 0 & \cdots\\
0 & 0 & \multicolumn{2}{c}{\multirow{2}{*}{\large $M_{2,i}$}}\\
\vdots & \vdots &  & \\
\end{array}\right]
\end{align}

To get $(D^T\Sigma^{-1}D)^{-1}$ we will start by inverting $\sum_{i=1}^n M_{1,i}$:
\begin{align}
(\sum_{i=1}^n M_{1,i})^{-1} &= \left(\sum_{i=1}^n \begin{bmatrix}
\dd_Z^T A_i^{-1}\dd_Z & \dd_Z^T A_i^{-1}\dd_{XZ}\\
\dd_{XZ}^T A_i^{-1}\dd_{Z} & \dd_{XZ}^T A_i^{-1}\dd_{XZ}
\end{bmatrix}\right)^{-1}\\
&=\left(\sum_{i=1}^n\begin{bmatrix}
\sum_{j=1}^T \frac{\sigma^2_{i,j}}{|A_{i,j}|} & \sum_{j=1}^T \frac{X_{i,j}\sigma^2_{i,j}}{|A_{i,j}|}\\ \sum_{j=1}^T \frac{X_{i,j}\sigma^2_{i,j}}{|A_{i,j}|}& \sum_{j=1}^T \frac{X_{i,j}\sigma^2_{i,j}}{|A_{i,j}|}
\end{bmatrix}\right)\\
&=\left(\sum_{i=1}^n\sum_{j=1}^T \frac{\sigma^2_{i,j}}{|A_{i,j}|}\begin{bmatrix}
1&X_{i,j}\\
X_{i,j}&X_{i,j}\\
\end{bmatrix}\right)^{-1}\label{eq:mi1_1}
\end{align}

We can simplify by considering the value of $|A_{i,j}|$ 
\begin{align}
|A_{i,j}|&=(x_{i,j,1}+x_{i,j,2})\sigma_{CT}^2+x_{i,j,1}x_{i,j,2}\\
&=\bb\frac{\sve}{m_{i,j}}\left(\frac{1}{\piz}+\frac{1}{1-\piz}\right)+\frac{(\sve)^2}{m_{i,j}^2\piz(1-\piz)}\\
&=\frac{\sve}{m_{i,j}\sigma_Z^2}\sigma_{CT}^2+\left(\frac{\sve}{m_{i,j}}\right)^2\frac{1}{\sigma_Z^2}\\
&=\frac{\sve}{m_{i,j}\sigma_Z^2}\left(\sigma_{CT}^2+\frac{\sve}{m_{i,j}}\right)\\
&=\frac{\sve\sigma_{i,j}^2}{m_{i,j}\sigma_Z^2}.
\end{align}
Substituting this into (\ref{eq:mi1_1}):
\begin{align}
(\sum_{i=1}^n M_{1,i})^{-1}&=\left(\sum_{i=1}^n\sum_{j=1}^T \frac{\sigma^2_{i,j}m_{i,j}\sigma_Z^2}{\sve\sigma_{i,j}^2}\begin{bmatrix}
1&X_{i,j}\\
X_{i,j}&X_{i,j}\\
\end{bmatrix}\right)^{-1}\\
&=\left(\sum_{i=1}^n\sum_{j=1}^T \frac{m_{i,j}\sigma_Z^2}{\sve}\begin{bmatrix}
1&X_{i,j}\\
X_{i,j}&X_{i,j}\\
\end{bmatrix}\right)^{-1}
\end{align}
Then defining 
\begin{align}
N &=  \sum_{i=1}^n\sum_{j=1}^T m_{i,j}\\
N_{X=1} &= \sum_{i=1}^n\sum_{j=1}^T X_{i,j}m_{i,j}\\
N_{X=0} &= \sum_{i=1}^n\sum_{j=1}^T (1-X_{i,j})m_{i,j},
\end{align}
\emph{i.e.}, the number of observations in total, the number of observations with $X=1$ and the number of observations with $X=0$, respectively, gives 
\begin{align}
(\sum_{i=1}^n M_{1,i})^{-1}&=\frac{\sve}{\sigma_Z^2}
\left(\begin{bmatrix}
N&N_{X=1}\\
N_{X=1}&N_{X=1}
\end{bmatrix}\right)^{-1}
\\
&=\frac{\sve}{\sigma_Z^2 N_{X=1}(N-N_{X=1})}\begin{bmatrix}
N_{X=1}&-N_{X=1}\\
-N_{X=1}&N
\end{bmatrix}
\\
&=\frac{\sve}{\sigma_Z^2 N_{X=0}}
\begin{bmatrix}
1&-1\\
-1&\frac{N}{N_{X=1}}
\end{bmatrix}\label{eq:mi1_2}
\end{align}

Then the variance of the individual and interaction treatment effect estimators are the (1,1) and (2,2) entries of (\ref{eq:mi1_2}), respectively. Substituting $\sve=(1-\rhoCT)\sigma^2$ we have
\begin{align}
\var(\hbetaI) &= \frac{(1-\rhoCT)\sigma^2}{\pi_Z(1-\pi_Z)N_{X=0}}\mbox{, and}\\
\var(\hbetaCI) &= \frac{(1-\rhoCT)\sigma^2 N}{\pi_Z(1-\pi_Z)N_{X=1}N_{X=0}}.
\end{align}
As in Appendix \ref{sec:2}, the marginal effect of the cluster-period-level intervention $\var(\hat{\tilde{\beta}}_2)$ is the same as we would get in an LCRT of the same design but with only a single-level (cluster-period-level) intervention. We can derive conditional values of this intervention effect in the same way as in Appendix \ref{sec:2}. Recall that $\betaC = \tbetaC-\pi_Z\betaCI$ so
\begin{align}
\var(\htbetaC) &= \VL \label{eq:beta2_var_int}\\
\var(\hbetaC) &= \var(\htbetaC-\pi_Z\hbetaCI) \label{eq:beta3_var_int}\\
&= \var(\htbetaC)+\pi_Z^2\var(\hbetaCI)-2\pi_Z^2\cov(\htbetaC,\hbetaCI) \\
&= \VL+\frac{(1-\rhoCT)\sigma^2 \pi_Z N}{(1-\pi_Z)N_{X=1}N_{X=0}}\label{eq:beta4_var_int}.
\end{align}
When all cluster period sizes are assumed to be equal, then $m_{i,j}=m$ for all $i,j$ and so $N_{X=1} = \pi_XmnT$ and $N= mnT$. In this case, all the variance expressions here reduce to the variance expressions given in Appendix \ref{sec:2} for the situation without variable cluster period sizes. 

\section{Differences from control-control condition}
We can also get variances where our effects of interest are mean differences from those with control-control. Call these differences $\boldsymbol{\beta}^*=(\beta_I^*,\beta_{IC}^*,\beta_I^*)^T$. Then:
\begin{align*}
\mathrm{Cov}(\hat{\boldsymbol{\beta}}^*) &= \begin{bmatrix} 
K\pix & K\pix\piz & K\pix\piz\\
K\pix\piz & K\pix\piz+V & K\pix(2\piz-1)+V\\
K\pix\piz & K\pix(2\piz-1)+V & K(\pix(2\piz-1)+(1-\piz)^2)+V
\end{bmatrix}
\end{align*}

\section{Variable cluster period sizes, no interaction term}\label{sec:var_size_no_int}
It is also possible to derive variance expressions when there is assumed to be no interaction between the different factors, in the case of varying cluster-period sizes. In this case there is no interaction term ($\betaCI$ or $\tbeta_4$) in the model, but otherwise the situation is identical to that outlined in Appendix \ref{sec:var_size}. Just as in Section \ref{sec:no_int} we just lose a row and column of $D^T\Sigma^{-1} D$. So now
\begin{align}
\left(\sum_{i=1}^n M_{1,i}\right)^{-1} &= \left(\sum_{i=1}^n\sum_{j=1}^T \frac{m_{i,j}\sigma_Z^2}{\sve}\right)^{-1} \\
&=\frac{(1-\rhoCT)\sigma^2}{\sigma_Z^2 N}
\end{align}
So 
\begin{align}
\var(\hbetaC) =  \VL\label{eq:beta2_var_no_int}\\
\var(\hbetaI) =\frac{(1-\rhoCT)\sigma^2}{\sigma_Z^2 N}\label{eq:beta3_var_no_int}
\end{align}

\end{document}